\documentclass[12pt]{article}
\catcode`\@=11
\@addtoreset{equation}{section}

\global\arraycolsep=2pt
\usepackage{mathrsfs,amsbsy,amssymb,latexsym,amsfonts,amsmath,cite}
\usepackage{graphicx,color}
\usepackage{physics}
\usepackage{mathtools}
\usepackage[normalem]{ulem}
\usepackage{caption}
\usepackage{here}

\usepackage{mathrsfs,amsbsy,amssymb,latexsym,amsfonts,amsmath,cite,bm}
\usepackage{graphicx,color}
\usepackage{mathtools}
\usepackage{enumerate}

\DeclareFontFamily{U}{BOONDOX-calo}{\skewchar\font=45 }
\DeclareFontShape{U}{BOONDOX-calo}{m}{n}{
  <-> s*[1.05] BOONDOX-r-calo}{}
\DeclareFontShape{U}{BOONDOX-calo}{b}{n}{
  <-> s*[1.05] BOONDOX-b-calo}{}
\DeclareMathAlphabet{\mathcalboondox}{U}{BOONDOX-calo}{m}{n}
\SetMathAlphabet{\mathcalboondox}{bold}{U}{BOONDOX-calo}{b}{n}
\DeclareMathAlphabet{\mathbcalboondox}{U}{BOONDOX-calo}{b}{n}

\allowdisplaybreaks

\begin{document}

\renewcommand{\thefootnote}{\fnsymbol{footnote}}

\begin{flushright}
STUPP-25-280
\end{flushright}
\vspace*{0.5cm}

\begin{center}
{\Large \bf  Resonances in Lifetimes of AdS Oscillon}
\vspace*{1.2cm} \\
{\large
Takaki Matsumoto$^{1}$\footnote{E-mail:~takaki-matsumoto$\_$at$\_$ejs.seikei.ac.jp},
Kanta Nakano$^{2}$\footnote{E-mail:~k.nakano.233$\_$at$\_$ms.saitama-u.ac.jp}, 
Ryosuke Suda$^{2}$\footnote{E-mail:~r.suda.813$\_$at$\_$ms.saitama-u.ac.jp} \\ 
and Kentaroh Yoshida$^{2}$\footnote{E-mail:~kenyoshida$\_$at$\_$mail.saitama-u.ac.jp}} 
\end{center}

\vspace*{0.4cm}

\begin{center}
$^{1}${\it Seikei University, 
3-3-1 Kichijoji-Kitamachi, Musashino-shi, Tokyo 180-8633, Japan}
\end{center}
\begin{center}
$^{2}${\it Graduate School of Science and Engineering, Saitama University, \\
255 Shimo-Okubo, Sakura-ku, Saitama 338-8570, Japan}
\end{center}

\vspace{1cm}

\begin{abstract}
Oscillons are classical oscillatory solutions with very long but finite lifetimes in real scalar field theories with appropriate potentials. An interesting feature is that resonances appear in the lifetimes of the oscillon for the initial size of the oscillon core $R_0$, which was discovered by Honda and Choptuik in the case of Minkowski space. In a previous work, oscillons in the global anti-de Sitter (AdS) space have been constructed, which we abbreviate as AdS oscillons. We present new resonance structures for the curvature radius $L$ and the core size $R_0$ in the lifetime of the AdS oscillon. We then compute exponents associated with the resonance peaks. Finally, we observe the bifurcation of the peaks due to the reflected waves.    
\end{abstract}

\setcounter{footnote}{0}
\setcounter{page}{0}
\thispagestyle{empty}

\newpage

\tableofcontents

\renewcommand\thefootnote{\arabic{footnote}}

\section{Introduction}

Oscillons are classical oscillatory solutions with longevity in real scalar field theories with appropriate potentials. These were originally discovered by Bogolubsky and Makhankov \cite{Bogolyubsky:1976nx,Makhankov:1978rg} and then their existence was confirmed by Gleiser \cite{Gleiser:1993pt}. The oscillons are not topological solitons because they have no topological charges. Furthermore, in contrast to Q-balls \cite{Coleman:1985ki}, there is no conserved charge as well. Therefore, oscillons are not completely stable and will decay despite having very long lifetimes.

\medskip 

A theoretical framework to explain the longevity of oscillons had been investigated in a series of papers \cite{Copeland:1995fq,Gleiser:2008ty,Gleiser:2009ys} and further elaborated to the present (For a recent summary of the progress, see \cite{Zhou:2024mea}). Currently, it is possible to easily construct oscillons under some suitable conditions phenomenologically supported, and it is expected to have applications in various fields including cosmology. However, the fundamental mechanism underlying their long lifetimes has yet to be elucidated, and it remains shrouded in mystery.

\medskip 

Recently, oscillons in the global anti-de Sitter (AdS) space have been studied in \cite{Ishii:2024yup}\footnote{For breather solution in AdS space, see \cite{Fodor:2013lza}.}. In the following, these solutions are abbreviated as AdS oscillons. The systems considered there are real scalar field theories with anti-symmetric double well potentials\footnote{The standard construction of oscillons includes a mass term. For a massless oscillon, see \cite{Dorey:2023sjh}.} in the global AdS space. 
The solutions are supposed to be spherically symmetric. In particular, the initial configuration is a Gaussian wave packet located at the origin of the global AdS space. In contrast to the Minkowski case \cite{Copeland:1995fq}, the presence of the finite curvature radius $L$ leads to two interesting features: 
\begin{align}
    \mbox{1) \quad the recurrence phenomenon},   \qquad 
    \mbox{2) \quad the existence of stable oscillating solutions}  \notag 
\end{align}

The first one is based on reflected waves. The oscillons emit radiations and collapse into a collection of small amplitude waves. Then their reflection occurs at a classical turning point. This is an advantage of the AdS space since no boundary conditions need to be imposed, in contrast to a finite box. Then the recurrence phenomenon occurs at the origin and the recurrent wave packets again show a noticeably long lifetime. With each reflection at a classical turning point, its lifetime becomes shorter and shorter. This is to be anticipated since the system is not integrable and perfect recurrence cannot be expected. However, we can still observe the recurrence phenomena with good accuracy.

\medskip 

The second is related to a finite curvature radius $L$\,. When the initial Gaussian core size $R_0$ is close to $L$\,, the time it takes for the reflected waves to return to the origin is significantly reduced. Then the reflected waves return to the origin during the decay of the oscillon. In other words, the decay of the oscillons is forbidden by the strong curvature effect of AdS space.

\medskip 

In this letter, we consider the third feature of the AdS oscillons: the resonance structure in lifetimes of AdS oscillons. The resonances with respect to the oscillon core size $R_0$ in lifetimes of flat space oscillons were originally discovered by Honda and Choptuik \cite{Honda:2001xg} and further elaborated by Gleiser and Krackow \cite{Gleiser:2019rvw}. In the AdS case, in addition to $R_0$\,, there exists the curvature radius $L$\,. Thus, revealing the resonance structure with respect to $L$ is an interesting issue. We present the resonance structure as a 3D plot for both $R_0$ and $L$\,. Then, we consider several sections and calculate the relevant exponents of some resonance curves.  Finally, we observe the bifurcation of the peaks due to the reflected waves.    

\medskip

This letter is organized as follows. In section 2, we introduce the setup for considering oscillons and explain some basic properties of AdS oscillons. In section 3, we show the resonance structure for both the core size $R_0$ and the curvature radius $L$\,. We then take several sections and calculate the exponents of some resonance curves. Finally, the bifurcation of the peaks is observed. Section 4 is devoted to the conclusion and discussion.

\section{Setup}

In this section, we present the setup of the AdS oscillons, following \cite{Ishii:2024yup}. 

\subsection{Classical action}

The background spacetime considered in this work is the $(3+1)$-dimensional global AdS space with radius of curvature $\ell$. The metric of this spacetime is written as
\begin{align}
    ds^2 = g_{\mu\nu} dx^{\mu} dx^{\nu}
    =
    -\left(1+\frac{r^2}{\ell^2}\right)dt^2
    +\left(1+\dfrac{r^2}{\ell^2}\right)^{-1}dr^2
    +r^2d\Omega_2^2\,,
\end{align}
where $r$ is the radial coordinate in three dimensions and $d\Omega_2^2$ denotes the round metric on the two-dimensional unit sphere. We study spherically symmetric solutions of a real scalar field theory in the AdS space described by the Lagrangian 
\begin{align}
    \mathcal{L}\left[\phi\right]
    = 
    4\pi r^2\left[
    -\frac{1}{2}\,g^{tt}\,\partial_{t}\phi\,\partial_{t}\phi 
    -\frac{1}{2}\,g^{rr}\,\partial_{r}\phi\,\partial_{r}\phi 
    - V(\phi) \right]\,.
    \label{action}
\end{align}
Here $V(\phi)$ is a symmetric double-well potential given by 
\begin{align}
    V(\phi) = \frac{\lambda}{4}\left(\phi^2 - \frac{m^2}{\lambda}\right)^2\,, 
    \label{symmetric_double_well}
\end{align}
with a mass parameter $m$ and a coupling $\lambda$.

\medskip

In the following, all quantities are made dimensionless using the mass parameter $m$. 
We then obtain the dimensionless Lagrangian, 
\begin{align}
    \mathcal{L}_\mathrm{rescaled}\left[\phi\right]
    =
    4\pi r^2
    \left[
    -\frac{1}{2f}\,
    \partial_t\phi\,\partial_t\phi
    -\frac{f}{2}\,
    \partial_r\phi\,\partial_r\phi
    -\frac{1}{4}(\phi^2 - 1)^2
    \right]\,,
    \label{Lagrngian_rescaled}
\end{align}
where we have set $\lambda=1$ and defined the function $f(r)$ by
\begin{align}
    f(r) \equiv 1 + \frac{r^2}{L^2}\,,\qquad L \equiv \ell\, m\,.
\end{align}
See Appendix A in the work \cite{Ishii:2024yup} for details of the rescaling.

\medskip 

For the dimensionless Lagrangian \eqref{Lagrngian_rescaled}, the equation of motion reads
\begin{align}
    -f^{-1}\,\partial_{t}^{2}\phi + f\,\partial_{r}^{2}\phi + \frac{2}{r}f\,\partial_{r}\phi +  \frac{2r}{L^2}\partial_{r}\phi
    - \phi(\phi^2-1) = 0\,. 
    \label{EoM_dimensionless}
\end{align}

\subsection{Oscillon ansatz and shell energy}

The AdS oscillons are solutions of the equation \eqref{EoM_dimensionless} satisfying the following four conditions:
\begin{align}
 &   \partial_t\phi(t=0,\,r) =0\,, \quad  
 \partial_r\phi(t,\,r=0) = 0\,, \quad \phi(t,\,r=\infty) = - 1\,,  \\ 
&    \phi(t=0,\,r) = 2{\rm e}^{-r^2/R_0^2} - 1\, . \label{bc3}
\end{align}
Note here that there are two minima $\phi=\pm 1$ in the (rescaled) symmetric double-well. The field approaches $-1$ as $r$ goes to infinity. The Gaussian configuration with the core size $R_0$ lies on the vacuum $\phi=-1$\,. 
Oscillons are the Gaussian wave packets which are localized and pinned at the origin $r=0$ and retain their shape for a very long but finite time. They emit radiation waves as time goes on and eventually decay.
 
\medskip

To monitor the time evolution of the oscillon and to detect radiations emitted from it, we introduce the shell energy as
\begin{align}
    E_{\rm s}(t) \equiv \int_0^{R_\mathrm{s}}\!\!dr\,\mathcal{E}\left[\phi\right]\,, 
    \label{Es}
\end{align}
where $R_{\rm s}$ is a suitably large radius compared to the initial core size $R_{0}$ such that a shell of radius $R_s$ encloses the initial Gaussian configuration. The integrand $\mathcal{E}[\phi]$ represents the energy density given by
\begin{align}
    \mathcal{E}\left[\phi\right]
    =
    4\pi r^2\left[\frac{1}{2f}\,\partial_{t}\phi\,\partial_{t}\phi 
    + \frac{f}{2}\,\partial_r\phi\,\partial_r\phi
    + \frac{1}{4}(\phi^2 - 1)^2
    \right]\,.
\end{align}
The shell energy contains all the energy of the oscillon at $t=0$ and decreases as the radiation waves escape from the shell. The total energy, which is the integral of $\mathcal{E}[\phi]$ over the entire space, is conserved.

\subsection{Conformal transformation}

It is helpful to convert the infinite radial direction $0\leq r <\infty$ to a finite interval $0\leq \theta <\pi/2$ 
by using the following conformal transformation, 
\begin{equation}
    r = L\,\tan{\theta}\,. 
\end{equation} 
As a result, the equation of motion is given by 
\begin{align}
    -\partial^2_{t}\phi
    +\frac{1}{L^2}\partial^2_\theta\phi
    +\frac{2}{L^2\sin{\theta}\cos{\theta}}\partial_\theta\phi
    -\frac{1}{\cos^2{\theta}}\phi(\phi^2-1) = 0\,.
\label{diff-dimless}
\end{align}
In the following, the original time $t$ is kept in comparison to the previous work \cite{Ishii:2024yup}. This is just because the computational cost is reduced. Similarly, the initial configuration \eqref{bc3} becomes 
\begin{align}
    \phi(t=0,\,r) = 2\exp\left(-\frac{L^2\tan^2{\theta}}{R_0^2}\right)-1\,.
\end{align}
In addition, the shell energy $E_{\rm s}$ is rewritten as  
\begin{align}
    E_{\rm s}(t) = 4\pi\,L^{3}\int_0^{\theta_\mathrm{s}}\!\!d\theta\,\,\tan^{2}\!{\theta}\left[\frac{1}{2}
    \partial_{t}\phi\partial_{t}\phi
    + \frac{1}{2L^2}
    \partial_\theta\phi\partial_\theta\phi
    + \frac{1}{4\cos^2{\theta}}\,(\phi^2 - 1)^2
    \right]\,, 
    \label{Es}
\end{align}
where $\theta_{\rm s} = \tan^{-1}\!\left({R_{\rm s}/L}\right)$\,. The total energy is obtained by taking that $\theta_{\rm s} = \pi/2$\,.

\section{Resonances in lifetimes of AdS oscillons} 

By performing numerical computations for the differential equation (\ref{diff-dimless})\,, oscillons can be constructed. The oscillons have very long but finite lifetimes. The decay process takes some time to complete, so we need to determine a working definition of oscillon's lifetime, where we say that the oscillon has decayed when its shell energy drops below half. 

\medskip 

Hereafter, we will compute the oscillon lifetime for some values of the oscillon core size $R_0$ and the curvature radius $L$\,. Honda and Choptuik computed lifetimes of oscillons in Minkowski spacetime by varying the values of the oscillon core size $R_0$ in \cite{Honda:2001xg}, where some coordinate transformations were performed to reduce the computational cost. We will not here perform the coordinate transformations but directly carry out the numerical analysis for the differential equation (\ref{diff-dimless})\,.   

\subsection{Resonances in the $R_0$-$L$ plane}

\begin{figure}[htbp]
    \centering
    \includegraphics[width=0.8\linewidth]{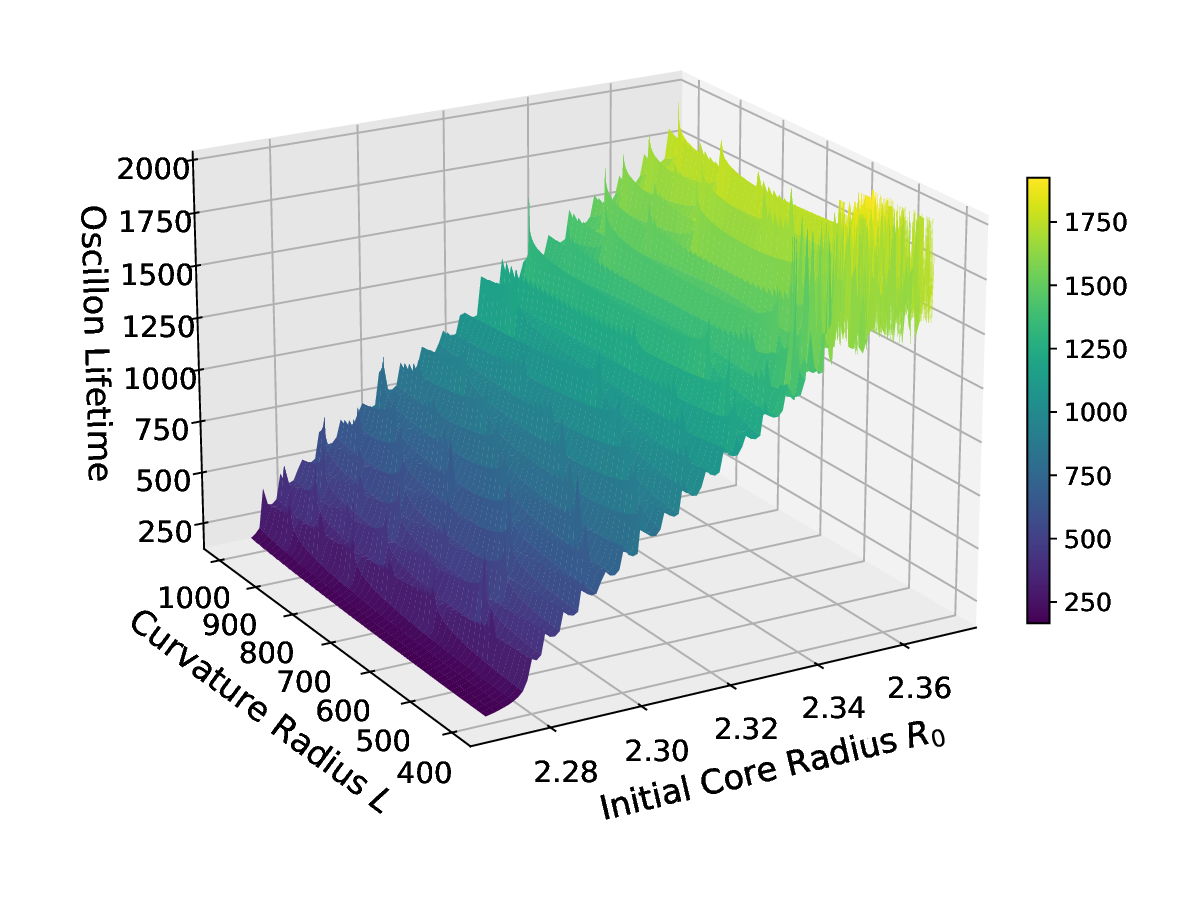}
    \vspace{-1cm}
    \caption{A plot of lifetimes of an AdS oscillon for $R_0$ and $L$.}
    \label{fig3-0}
\end{figure}

First of all, we shall see the global structure of the AdS oscillon lifetimes by plotting the lifetimes versus both $R_0$ and $L$\,. This is presented in Fig.\,\ref{fig3-0} as a 3D plot. The vertical axis is the lifetime and the horizontal axes are $R_0$ and $L$\,. One can clearly see a lot of peaks. By taking a section for a fixed value of $L$\,, one can see resonances for values of $R_0$ as in \cite{Honda:2001xg}. Then, by changing the value of $L$\,, the peaks shift to form ridges. Remarkably, new resonances appear along the ridges depending on the value of $L$\,.  

\medskip 

\begin{figure}[htbp]
    \centering
    \includegraphics[width=0.8\linewidth]{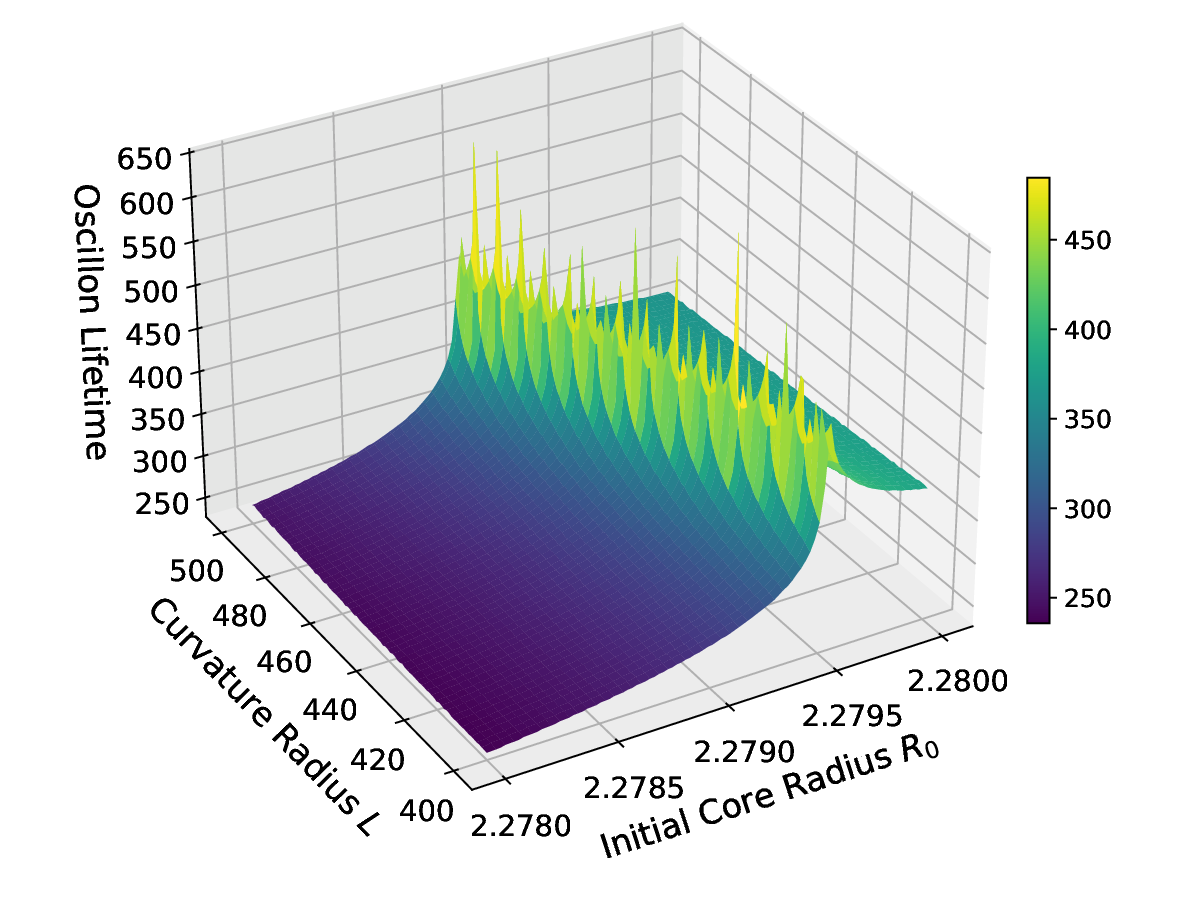}
    \caption{A zoom-up of the region with $2.278 \leq R_0 \leq 2.280$ and $400 \leq L \leq 500$ in Fig.\,\ref{fig3-0}.}
    \label{fig3-1}
\end{figure}

As in \cite{Honda:2001xg}, the resonances in the AdS oscillon lifetimes also exhibit the self-similar structure. Figure \ref{fig3-1} shows the zoom-up of the region with $2.278 \leq R_0 \leq 2.280$ and $400 \leq L \leq 500$ of Fig.\,\ref{fig3-0}. One can see a lot of small peaks and the figure exhibits a self-similar structure. This self-similarity along the $R_0$-direction was found in \cite{Honda:2001xg}. A remarkable point here is that a self-similar structure can also be seen along the $L$-axis as well.

\medskip 

In the following, let us see the detailed properties of peaks by considering sections of Fig.\,\ref{fig3-0}\,.

\subsection{Peaks and exponents for $R_0$ with fixed $L$} 

\begin{figure}[htbp]
    \centering
    \includegraphics[width=0.6\linewidth]{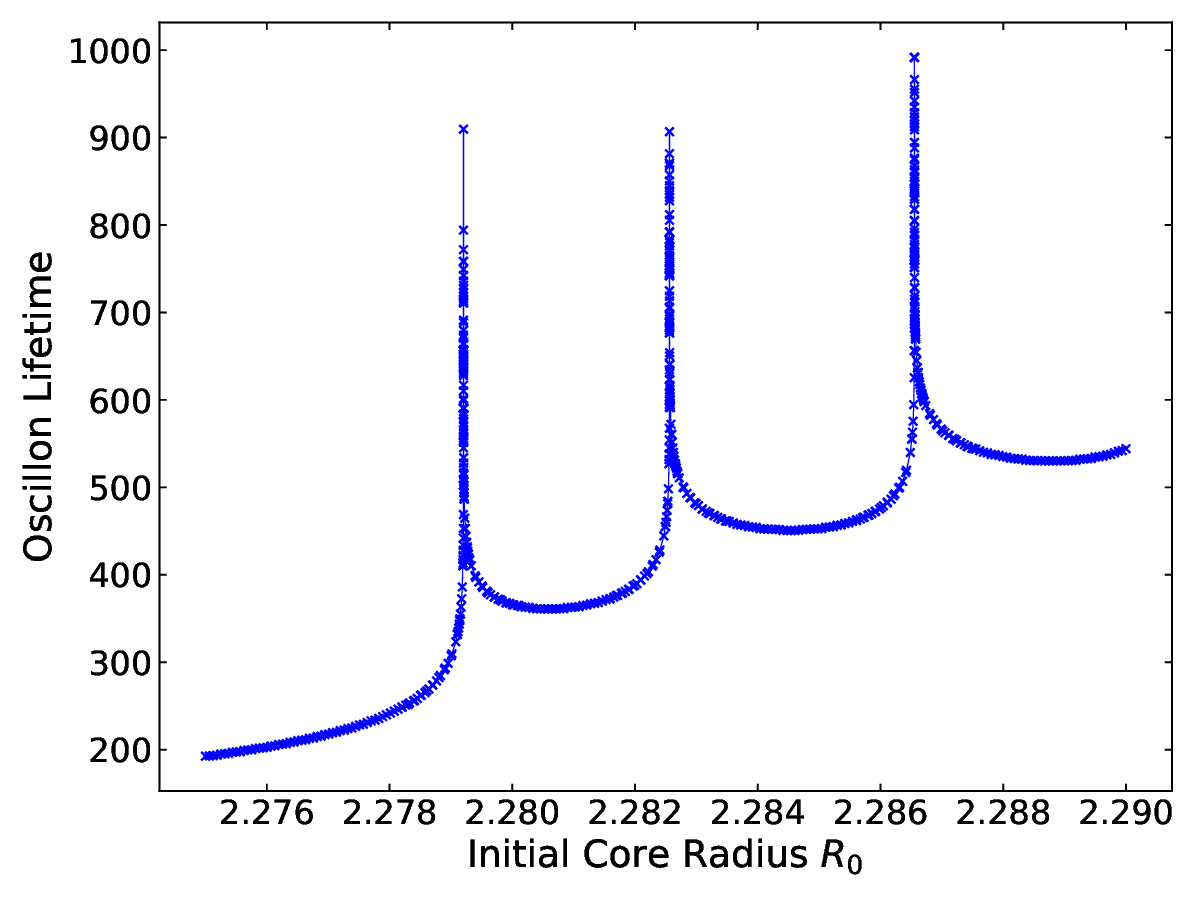}
    \caption{Lifetimes of the AdS oscillon for $R_0$ (fixed $L=500$).}
    \label{fig3}
\end{figure}

Let us consider here a section with fixed $L$\,. Figure \ref{fig3} shows a section with $L=500$ and lifetimes of the AdS oscillon are plotted for values of the core size $R_0$\,. There are three peaks in Fig.\,\ref{fig3}.  

\subsubsection*{Exponents} 

\begin{figure}[htbp]
    \centering
    \includegraphics[width=0.6\linewidth]{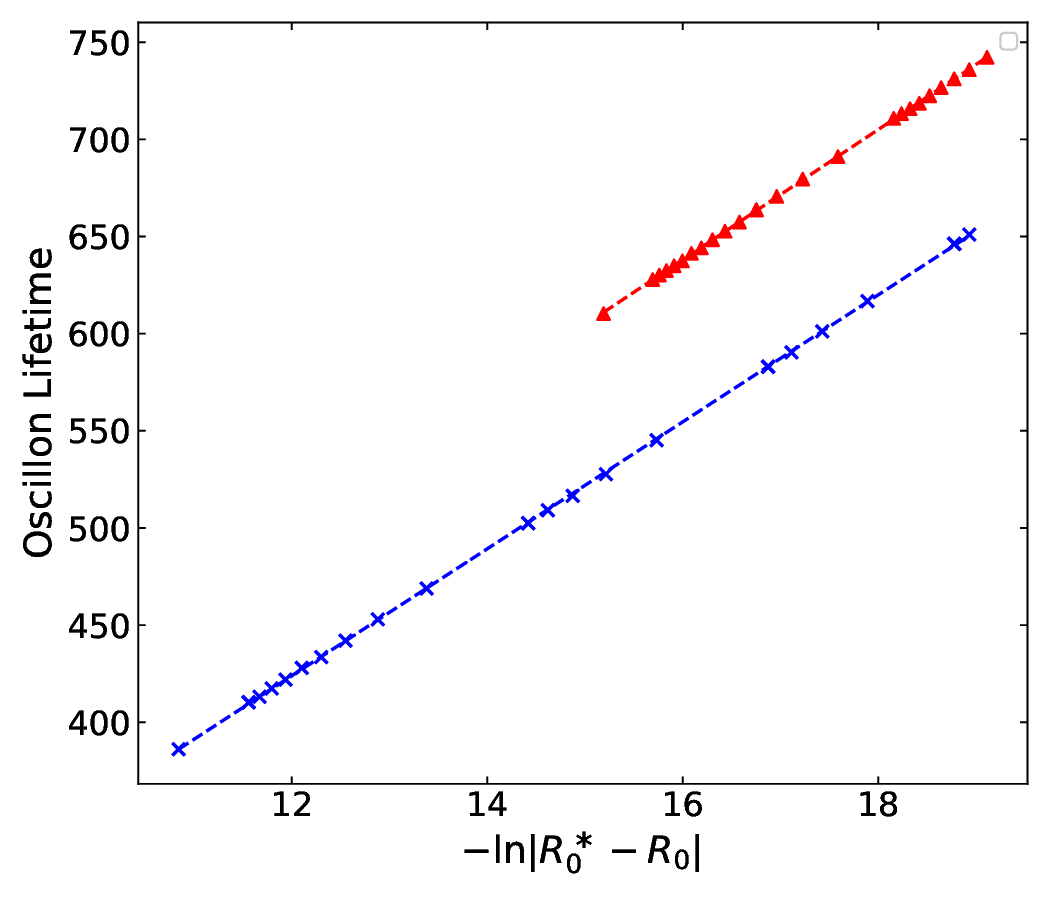}
    \caption{The exponents $\gamma_{\pm}$ of the peak with $R_0^{\ast} \approx 2.279$. The red (blue) marks are related to $\gamma_+ ~(\gamma_-)$\,.}
    \label{fig4}
\end{figure}

As in \cite{Honda:2001xg}, for each of the peaks, one can define the exponents $\gamma_{\pm}$ for the left and right curves, respectively, as 
\begin{eqnarray}
    \text{lifetime}
    ~=~ 
    \left\{
    \begin{array}{ll} 
    -\gamma_{+}\,\ln (R_0 - R_0^\ast) + \mbox{const.} & \quad 
    \mbox{(the right curve)}\\ 
    -\gamma_{-}\,\ln(R_0^{\ast}-R_0) + \mbox{const.} & \quad 
    \mbox{(the left curve)}
    \end{array}
    \right. \,,
\end{eqnarray}
where $R_0^{\ast}$ is the location of the peak on the $R_0$ line. The exponents can be numerically computed. Figure \ref{fig4} shows the result for the peak with $R_0^{\ast}\approx2.279$\,. For the other peaks, the exponents can be computed similarly. The resulting exponents for the three peaks are summarized as 


\begin{equation}
\begin{array}{c|ccc} 
~~ R_{0}^{\ast} ~~ & 2.279  &  2.283 & 2.287  \\
\hline 
\gamma_+ & ~~33.486~~ & ~~31.606~~ & ~~30.918 ~~ \\ 
\gamma_- & 32.673 & 33.876 & 33.725
\end{array}
\,. 
\end{equation}


\subsection{Peaks and exponents for $L$ with fixed $R_0$} 

\begin{figure}[htbp]
    \centering
    \includegraphics[width=0.6\linewidth]{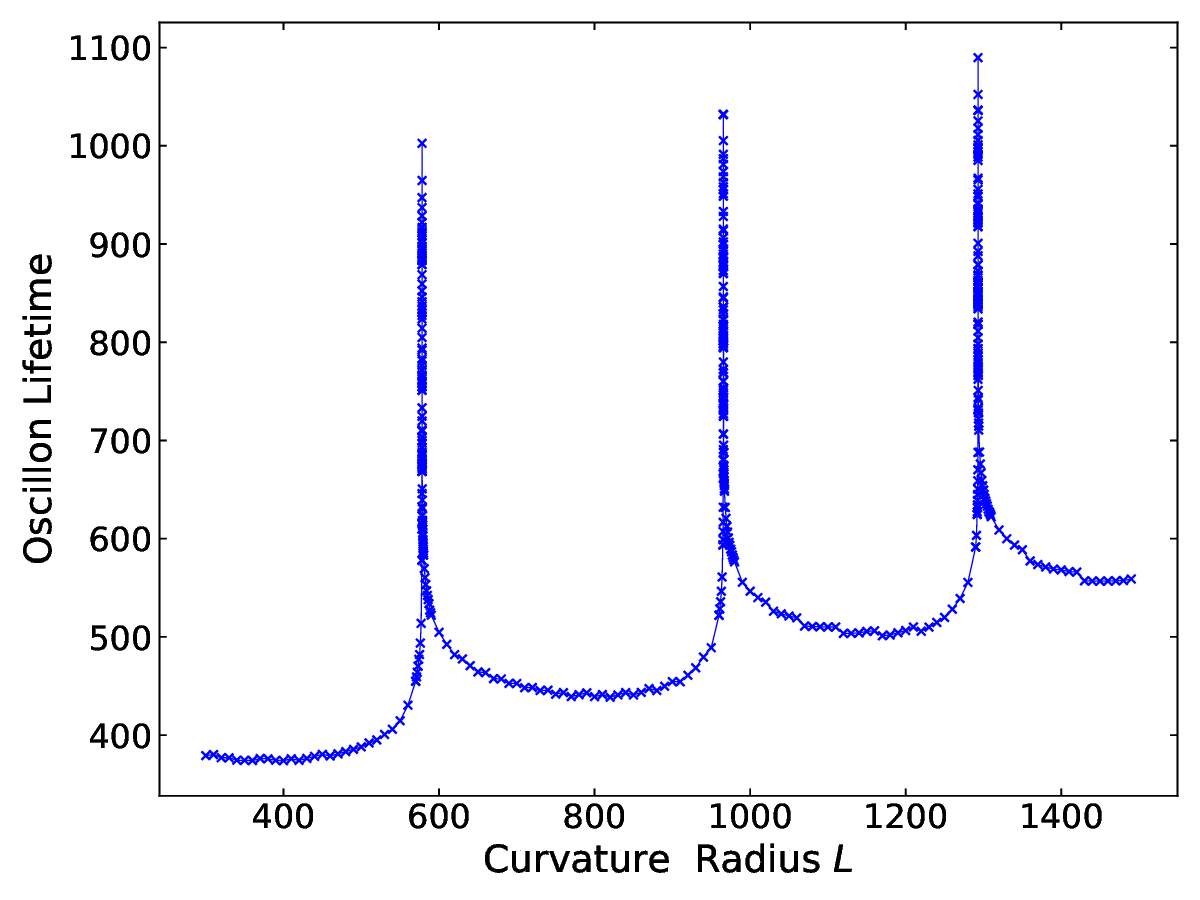}
    \caption{Lifetimes of the AdS oscillon for $L$ (fixed $R_0=2.282$).}
    \label{fig5}
\end{figure}

In this section, we consider a section with fixed $R_0$\,. Figure \ref{fig5} shows a section at $R_0=2.282$ and plots the lifetime of the AdS oscillon for values of AdS radius $L$. As plotted in Fig.\ref{fig5}, we find resonance peaks along the $L$-direction for the AdS oscillon lifetimes.

\subsubsection*{Exponents} 

\begin{figure}[htbp]
    \centering
    \includegraphics[width=0.6\linewidth]{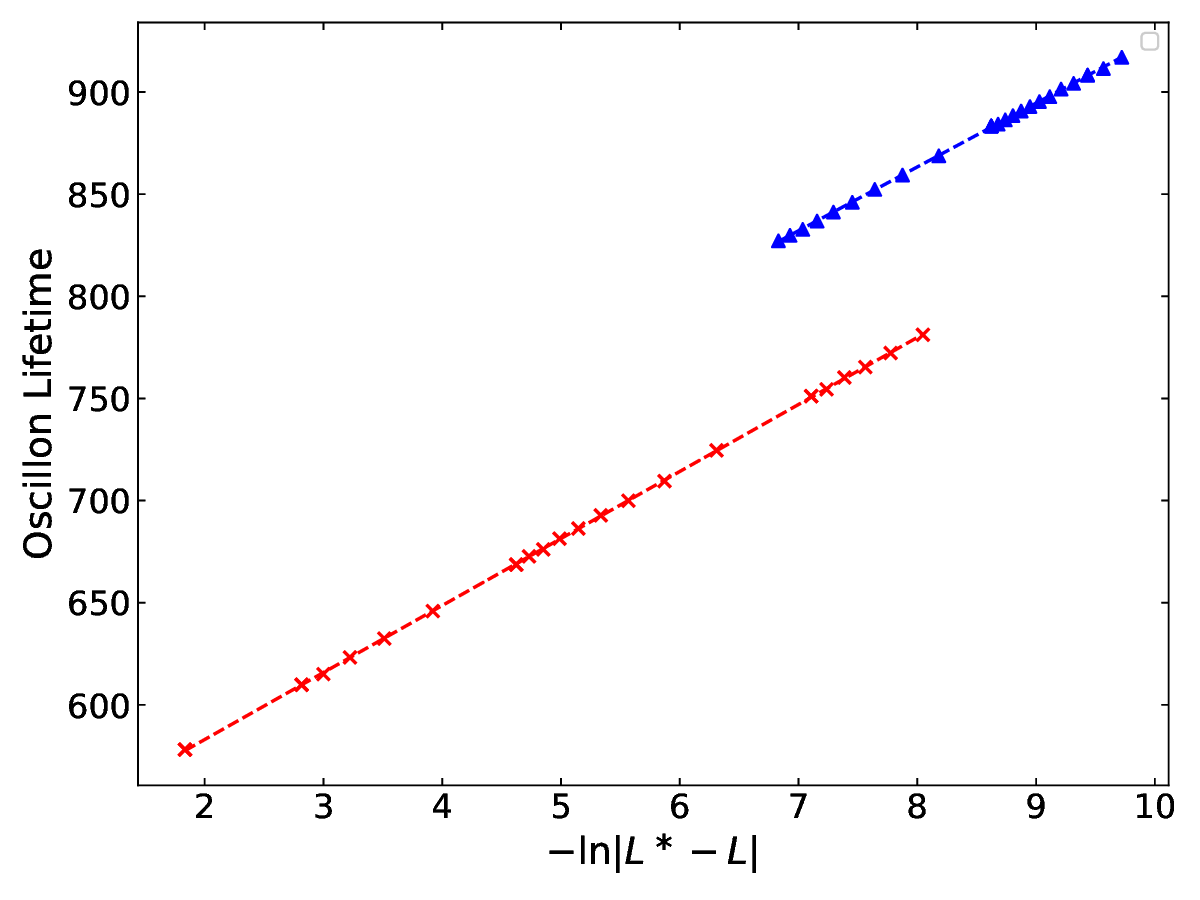}
    \caption{The exponents $\chi_{\pm}$ of the peak with $L^\ast \approx 578.160$. The red (blue) mark are concerned with values of $\chi_{+} ~(\chi_{-})$\,.}
    \label{fig6}
\end{figure}

For the peaks in the $L$-direction, we introduce new exponents $\chi_{\pm}$ for the two branches of the curve, assigning $\chi_{-}$ to the left curve and  $\chi_{+}$ to the right curve, as follows:
\begin{eqnarray}
    \text{lifetime}
    ~=~ 
    \left\{
    \begin{array}{ll} 
    -\chi_{+}\,\ln (L - L^\ast) + \mbox{const.} & \quad 
    \mbox{(the right curve)}\\  
    -\chi_{-}\,\ln(L^{\ast}-L) + \mbox{const.} & \quad 
    \mbox{(the left curve)}
    \end{array}
    \right. \,.
\end{eqnarray}
Here, $L^\ast$ denotes the location of a peak. In Fig.\,\ref{fig6}, the exponents are computed from the lifetimes around the peak at $L^\ast \approx 578.160$. Performing the same calculations for the other peaks, we obtain
\begin{equation}
\begin{array}{c|ccc} 
~~ L^{\ast} ~~ & 578.160  &  966.567 & 1293.032  \\
\hline 
\chi_+ & ~~32.803~~ & ~~32.228~~ & ~~31.065~~ \\ 
\chi_- & 31.217 & 30.395 &  30.089 
\end{array}
\,.
\end{equation}

\subsection{The bifurcation of peaks by reflected waves}

\begin{figure}[htbp]
    \centering
    \includegraphics[width=0.6\linewidth]{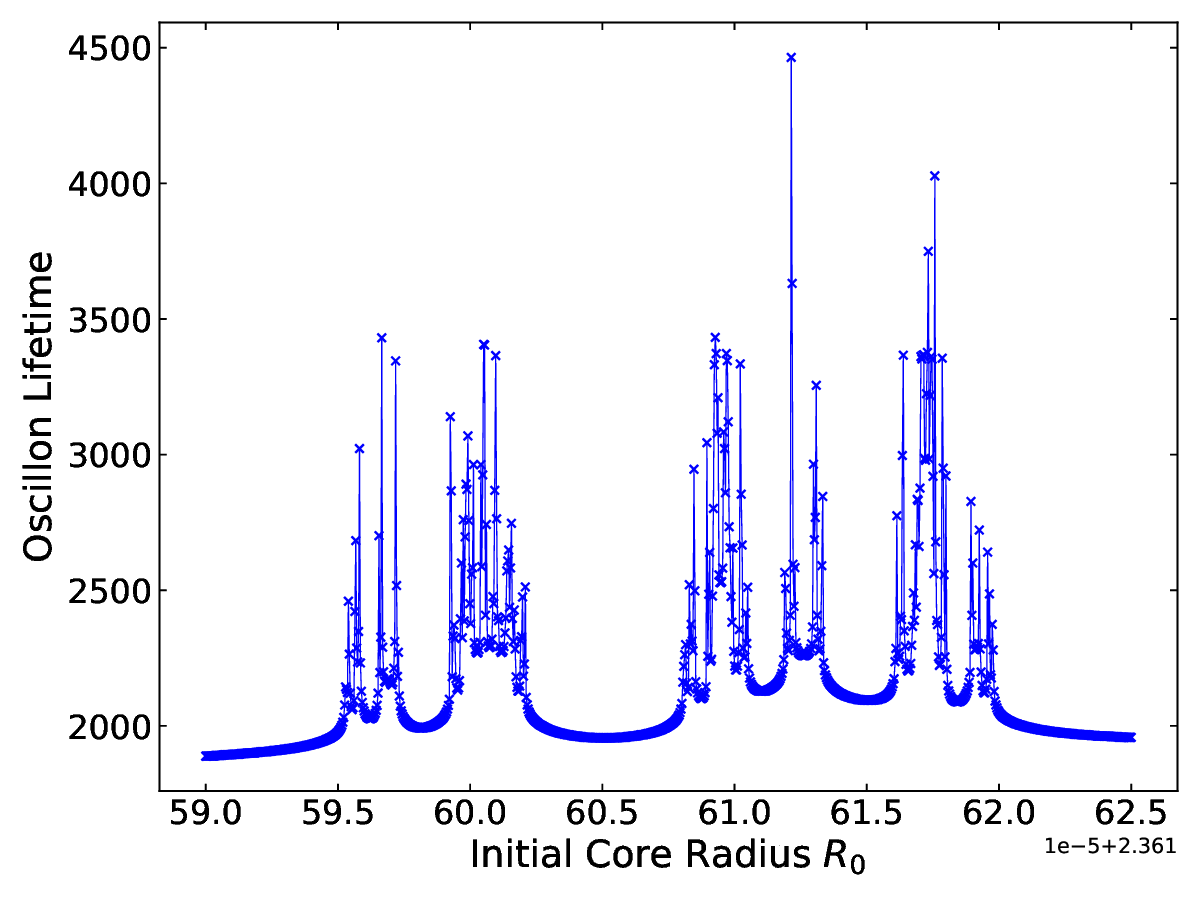}
    \caption{Resonance structure with reflected waves  for $R_0$ (fixed $L=500$).}
    \label{fig7}
\end{figure}

\begin{figure}[htbp]
    \centering
    \includegraphics[width=0.6\linewidth]{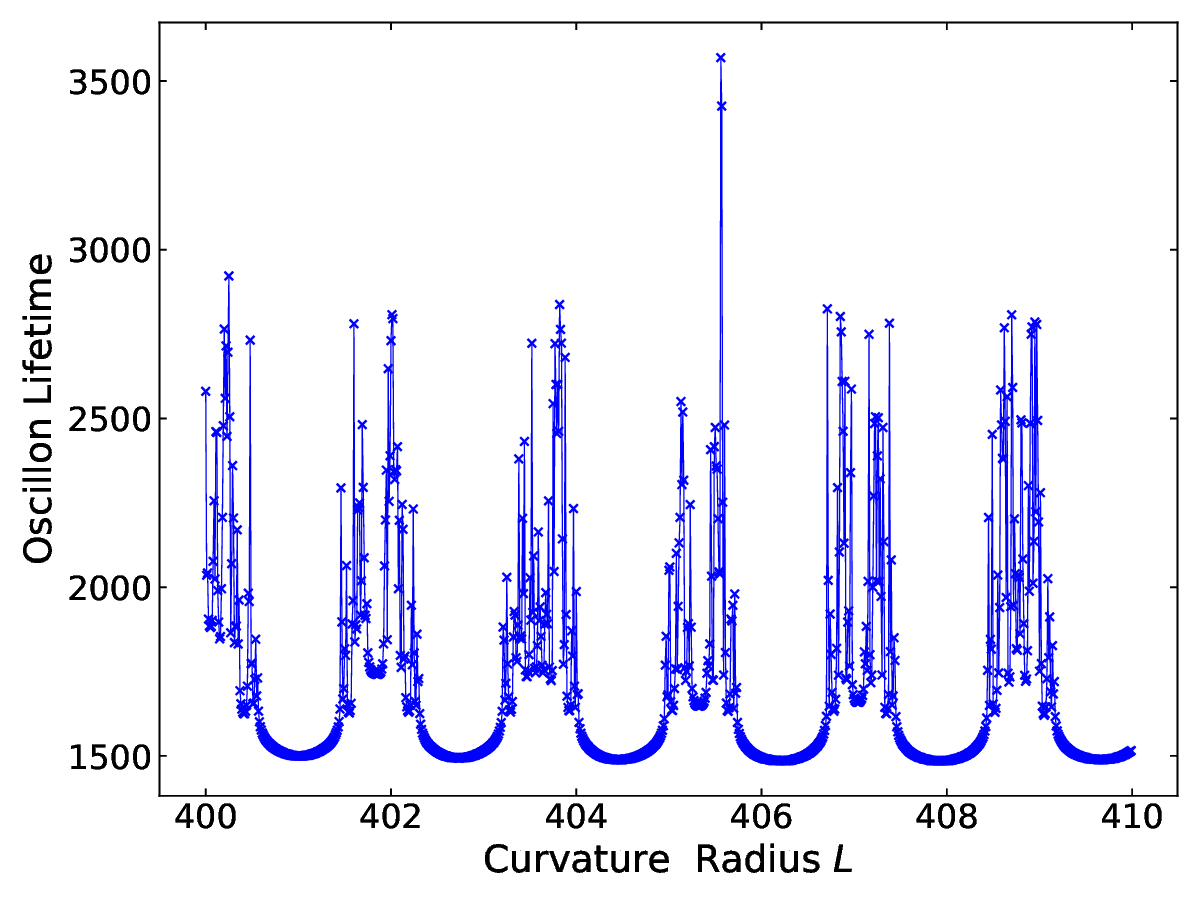}
    \caption{Resonance structure with reflected waves  for $L$ (fixed $R_0=2.36$).}
    \label{fig8}
\end{figure}

Finally, we shall see the effect of the reflected waves in the resonance structures. Figures \ref{fig7} and \ref{fig8} show lifetimes of the AdS oscillons for values of $R_0$ with fixed $L=500$ and for values of $L$ with fixed $R_0=2.36$, respectively. 
In both cases, the parameters are chosen to enhance the effect of the reflected waves. As a result, one can see the bifurcation of the resonance peaks. These bifurcations appear to follow a particular pattern, probably related to chaotic scattering \cite{chaotic}, in which a fractal structure emerges in the lifetimes\footnote{For the fractal structure in simple models, see \cite{Fukushima:2022lsd,Fukushima:2024agd}.}.



\section{Conclusion and Discussion} 

In this letter, we have continued to study oscillons in AdS space and found new resonance structures for the curvature radius $L$ and the oscillon core size $R_0$ in lifetimes of the AdS oscillon. We then have computed the exponents associated with the resonance peaks by taking sections with fixed $L$ or $R_0$\,. Finally, we have observed the bifurcation of the resonance  peaks due to the reflected waves. It seems likely that the bifurcation has some pattern. It would be interesting to study it more carefully.

\medskip 

There are some future directions. It is desirable to understand the exponents of the peak curves analytically. For this issue, it may be interesting to consider the simplest oscillon constructed by Manton and Roma \cite{Manton:2023mdr}. This oscillon was constructed in a real scalar field theory with a cubic potential in two-dimensional Minkowski spacetime and can be seen as a perturbative expansion around a sphaleron \cite{Klinkhamer:1984di}. Hence one may get some analytical understanding of the exponents. The oscillon in \cite{Manton:2023mdr} is constructed in two-dimensional Minkowski spacetime only, at least so far. So it is also interesting to construct the simplest oscillon in two-dimensional AdS space. 

\medskip 

As an application, it is intriguing to consider a holographic interpretation of the resonances discussed here via the AdS/CFT correspondence \cite{Maldacena:1997re}. In particular, in the AdS/QCD scenario \cite{Gursoy:2007cb,Gursoy:2007er} the dilaton may exhibit  fluctuations with longevity like oscillons before falling down to the stable configuration. The resonance structures may be related to some hadronic structures.  

\medskip 

It is also interesting to include the gravitational back reaction by following \cite{Bizon:2011gg}. The resonance structures may be bifurcated or destroyed by it. 

\medskip 

We hope that the resonance structures found here would shed light on new aspects of the AdS/CFT correspondence.

\subsection*{Acknowledgments}

The authors thank Takaaki Ishii, Tomohiro Shigemura and Norihiro Tanahashi for useful discussions and comments. The works of K.~Y.\ were supported by MEXT KAKENHI Grant-in-Aid for Transformative Research Areas A ``Machine Learning Physics'' No.\,22H05115, and JSPS Grant-in-Aid for Scientific Research (B) 
No.\,22H01217 and (C) No.\,25K07313.

\end{document}